# High-subwavelength-resolution imaging of multilayered structures consisting of alternating negative-permittivty and dielectric layers with flattened transmission curves

Yi Jin[1] and Sailing He[1,2,*]

[1]*Centre for Optical and Electromagnetic Research, Zhejiang University; Joint Research Centre of Photonics of the Royal Institute of Technology (Sweden) and Zhejiang University, Hangzhou 310027, China*

[2]*Division of Electromagnetic Engineering, School of Electrical Engineering, Royal Institute of Technology, S-100 44 Stockholm, Sweden*

[*]*Corresponding author: sailing@kth.se*

Multilayered structures consisting of alternating negative-permittivity and dielectric layers are explored to obtain high-resolution imaging of subwavelength objects. The peaks with the smallest $|k_y|$ ($k_y$ is the transverse wave vector) on the transmission curves, which come from the guided modes of the multilayered structures, can not be completely damped by material loss. This makes the amplitudes of the evanescent waves around these peaks inappropriate after transmitted through the imaging structures, and the imaging quality is not good. To solve such a problem, the permittivity of the dielectric layers is appropriately chosen to make these sharp peaks merge with their neighboring peaks. Wide flat upheavals are then generated on the transmission curves so that evanescent waves in a large range are transmitted through the structures with appropriate amplitudes. In addition, it is found that the sharp peaks with the smallest $|k_y|$ can be eliminated by adding appropriate coating layers and wide flat upheavals can also be obtained. © 2008 Optical Society of America

## 1. INTRODUCTION

Pendry suggested that a layer of negative index material (NIM) with $\varepsilon = -1$ and $\mu = -1$ can act as a perfect lens to restore all the information of an object at the image plane, including all the propagation waves and evanescent waves [1]. However, there are no natural NIMs and low-loss isotropic artificial NIMs are difficulty to fabricate, especially at the infrared and visible frequencies. As a reduced (easy for realization) version of a perfect lens, Pendry proposed that a thin noble metal film can be used as a super-lens to beat the diffraction limit and obtain subwavelength-resolution imaging by coupling incident evanescent waves into surface plasmon resonance and magnifying them [1]. The breakthrough of beating the diffraction limit with thin silver films has been experimentally verified [2]. Except noble metals, some other polar and semiconductor materials (i.e., SiC) can also possess negative permittivity and support surface plasmons at some optical frequencies. Some subwavelength imaging experiments using SiC have been reported [3,4]. Recently, subwavelength imaging has been extended from single negative-permittivity layers to multilayered structures consisting of alternating negative-permittivity layers and dielectric layers [5–15]. To reduce the influence of material loss from noble metals, Ramakrishna et al. cut a thin metal film into very thin layers and alternating them with dielectric layers [5,6]. These metal and dielectric layers possess the same thickness, and the real parts of their permittivities have the opposite signs. Another advantage of multilayered structures is that the real part of the negative permittivity is not required to be opposite to the permittivity of the environment while this is required for a single negative-permittivity super-lens. This is very beneficial for fabrication flexibility and having more freedom in choosing working frequencies. The structures constructed in Refs. 5 and 6 are a special case of so-called structures of canalization subwavelength imaging [7] which have very straight equifrequency contours (EFCs) parallel to the interfaces of the structures and transmit electromagnetic waves within the structures without the usual diffraction. Blomer at al. used the conception of coating layers to obtain broadband super-resolving Ag/GaP multilayered structures with high transparency in the visible range [13]. Shin et al. used metallodielectric stacks to realize all-angle negative refraction and obtain subwavelength imaging with excited waveguide modes magnifying the near field [8]. Waveguide modes in multilayered structures come from the coupling of plasmon waves on the interfaces of negative-permittivity layers [16]. Negative refraction of multilayered structures was also studied in [12], where far-field imaging without evanescent waves involved was also investigated. In the present paper, we consider the positive and negative influence of guided modes on the subwavelength imaging of multilayered structures. When guided modes, especially those with the smallest $|k_y|$ ($k_y$ is the transverse wave vector) exists, the incident evanescent waves are usually over-magnified. Over-magnification makes the amplitude proportion of the transmitted evanescent waves inappropriate and then degrades the subwavelength imaging. In this paper, by choosing appropriately the permittivity of dielectric layers according to the chosen negative permittivity, the guided modes with the smallest $|k_y|$, not very near $\pm k_0$ ($k_0$ is the wave number in free space), are nearly (but not yet) excited, and the incident evanescent waves in a large range are transmitted to the image plane with appropriate amplitudes so that high-subwavelength-resoluion imaging is obtained when material loss is considered. It is also found that adding appropriate coating layers on one side or both sides of multilayered structures can also give a similar effect.

This paper is organized as follows. In section 2, multilayered structures are first homogenized with the effective medium theory (EMT) [17]. The obtained effective homogeneous anisotropic layers (EHALs) are analyzed for subwavelength imaging. The incident evanescent waves in a large range are transmitted to the image plane with appropriate amplitudes using the suggested methods. In section 3, rigorous numerical simulation using the standard transfer-matrix method verifies the methods suggested in section 2 for the subwavelength imaging of some multilayered structures.

## 2. ANALYSIS OF MULTILAYERED STRUCTURES FOR HIGH-SUBWAVELENGTH-RESOLUTION IMAGING BASED ON THE EMT

A multilayered structure composed of alternating negative-permittivity layers and positive-permittivity dielectric layers is shown in Fig. 1. The relative permittivity and thickness of the negative-permittivity layers are denoted by $\varepsilon_1$ and $d_1$, respectively, and $\varepsilon_2$ and $d_2$ are for the dielectric layers. In this paper, when imaging is carried out, a line source as an object is put on the left surface (but in air) of the imaging structure, and the image plane is defined on the right surface (but in air). That is, the system is for near-field imaging. In the present paper, it is assumed that the magnetic field is perpendicular to the $x$-$y$ plane (TM polarization) and the time harmonic factor is $\exp(-i\omega t)$.

In subwavelength imaging of multilayered structures, the period $a=d_1+d_2$ is usually quite small compared with the wavelength $\lambda$. Thus, the multilayered structure is usually treated as an EHAL with the EMT, and the effective relative permittivity tensor is

$$\bar{\bar{\varepsilon}} = \varepsilon_x \mathbf{xx} + \varepsilon_y \mathbf{yy},  \tag{1}$$

where $\varepsilon_x = a/(d_1/\varepsilon_1 + d_2/\varepsilon_2)$ and $\varepsilon_y = (\varepsilon_1 d_1 + \varepsilon_2 d_2)/a$. The dispersion equation of the EHAL is as follows,

$$\frac{k_x^2}{\varepsilon_y} + \frac{k_y^2}{\varepsilon_x} = k_0^2. \tag{2}$$

In Eq. (2), one can see that if $\varepsilon_x$ is infinite, $k_x$ is $\pm\sqrt{\varepsilon_y}k_0$ independent of the value of $k_y$, which means that the EFCs of the EHAL are two straight lines parallel to the $y$ axis when $\varepsilon_y$ is positive.

Imaging of an EHAL in air with $\varepsilon_x = \infty$ and $\varepsilon_y = 0$ is investigated. Latter it will be explained why this special case is treated as a starting point to achieve high imaging resolution. All the multilayered structures investigated in [5,6] belong to this kind when they are homogenized. The transmission coefficient $t$ of the EHAL with thickness $d$ is as follows,

$$t(k_y) = \frac{2}{2\cos(k_x d) - i\left(\dfrac{k_x}{k_{0,x}\varepsilon_y} + \dfrac{k_{0,x}\varepsilon_y}{k_x}\right)\sin(k_x d)} \xrightarrow[\varepsilon_x=\infty,\varepsilon_y=0]{} \frac{2}{2 - i\dfrac{dk_0^2}{k_{0,x}}}, \tag{3}$$

where $k_{0,x} = \sqrt{k_0^2 - k_y^2}$. According to Eq. (3), one knows that $|t|$ is about 1 for $|k_y \Box\ k_0$ independent of $d$. This means that the evanescent waves with large $|k_y|$ can be transmitted through the EHAL nearly without attenuation or magnification. In Eq. (3), one can also know that the transmission coefficient $t$ is infinite for $k_y = \pm k_0\sqrt{1 + (dk_0)^2/4}$, which means that two, and only two, guided modes can exist in the EHAL. When *d* increases, the absolute values |$k_y$| of the transverse vectors of the two guided modes become larger. |*t*| is much larger than 1 in a range around $k_y = \pm k_0 \sqrt{1+(dk_0)^2/4}$, and the incident evanescent waves in this range are over-magnified after transmitted through the layer. For subwavelength imaging, one wishes to restore the original amplitudes of evanescent waves emitted from the object at the image plane. If perfect restoration is not possible, the amplitudes of some evanescent waves should not be over-magnified, otherwise, strong side-lobes may appear and destroy the imaging. Thus, the existence of sharp transmission peaks due to the excitation of guided modes is harmful for subwavelength imaging. Thus, the lossless EHAL with $\varepsilon_x = \infty$ and $\varepsilon_y = 0$ can not work well for subwavelength imaging even if such a lossless layer exists. In practice, material loss always exists in a natural negative-permittivity material. To investigate the influence of material loss on the above EHAL obtained from the homogenization of a multilayered structure, we choose the parameters of the corresponding multilayered structure as $\varepsilon_1 = -\varepsilon_2 = -3.5$, $d_1 = d_2 = a/2$, $d=9a$ and $a=\lambda/20$ before material loss is introduced. When the imaginary part of $\varepsilon_1$ is set to 0.23 ($\varepsilon_1 = -3.5+0.23i$ is the permittivity of SiC at $\lambda = 10.939 \, \mu m$ [3]), according to Eq. (1) the effective permittivity parameter ($\varepsilon_x$, $\varepsilon_y$) of the EHAL is (7.0 +106.52*i*, 0.115*i*). Curve 1 in Fig. 2(a) and curve 1 in Fig. 2(c) are the transmission curves of the EHAL corresponding to the lossless and lossy cases, respectively. Since the transmission |*t*| for negative $k_y$ is the same as that for positive $k_y$, we show |*t*| only for positive $k_y$ for all the transmission curves in this paper. With loss introduced, the transmission |*t*| drops quickly for large |$k_y$|, completely different from the lossless case. Fortunately such dropping of |*t*| for large |$k_y$| does not influence significantly high-resolution imaging. Among various wave components emitted from a subwavelength object, the amplitudes of evanescent waves with small |$k_y$| are usually larger than those with large |$k_y$|. For example, the spatial spectrum of a line source is as follows [18],

$$f(k_y) = \frac{1}{\sqrt{k_0^2 - k_y^2}}. \tag{4}$$

In Eq. (4), one can see that $f$ decreases nearly linearly as |$k_y$| increases for large |$k_y$|. That is, evanescent waves with small |$k_y$| are more important to generate a subwavelength image, and evanescent waves with large |$k_y$| are of little importance. Material loss damps the sharp transmission peaks, but they are still a bit sharp and over-magnification of some evanescent waves can not be eliminated completely. Curve 1 in Fig. 2(d) shows the distribution of magnetic field intensity at the image plane when material loss exists. The imaging quality is not good. Based on the above analysis, to obtain high imaging resolution, the key is to transmit evanescent waves in a wide range through the lens structure in appropriate proportion. Such a goal will be realized below.

We first investigate how the guided modes of a lossless EHAL are influenced by increasing the positive permittivity $\varepsilon_2$ of the corresponding multilayered structure. We use the same parameters as before, i.e., $\varepsilon_1 = -\varepsilon_2 = -3.5$, $d_1 = d_2 = a/2$, $d = 9a$ and $a = \lambda/20$. When $\varepsilon_2$ increases from 3.5 to 4.0, 4.3, 4.8 and 6.0, the effective permittivity parameter ($\varepsilon_x$, $\varepsilon_y$) of the EHAL changes from ($\infty$, 0) to (−56.0, 0.25), (−37.625, 0.4), (−25.8462, 0.65) and (−16.8, 1.25), and Fig. 2(a) shows the corresponding transmission curves. As shown in Fig. 2(a), with $\varepsilon_2$ increasing and deviating away from 3.5, many, not just one, peaks appear, and the leftmost peak moves toward the right while the other peaks move toward the left. With the two leftmost peaks getting closer, they merge together (both peaks disappear) and become a wide flat upheaval, as shown by curve 4 in Fig. 2(a). The merging point is not very near $k_0$ so that the wide flat upheaval can be obtained. The flat upheaval is very beneficial for subwavelength imaging. As $\varepsilon_2$ increases further, the upheaval contracts gradually to disappear and then a new sharp peak is generated very near $k_0$ as shown by curve 5 in Fig. 2(a). Then this new peak moves toward the right and merges with the nearest peak which keeps moving toward the left. The phenomenon of moving toward the left of the transmission peaks except the leftmost peak as $\varepsilon_2$ increases can be understood according to the change of the EFCs of the EHAL. As shown in Fig. 2(b), when increasing $\varepsilon_2$ from 3.5 to 4.0, 4.3, 4.8 and 6.0, the EFCs become hyperbolic and contract gradually. The hyperbolic shape leads to the appearance of many peaks, and the contraction of the EFCs forces the transmission peaks with large $|k_y|$ to move toward $\pm k_0$ as $\varepsilon_2$ increases. Then we introduce some material loss and set the imaginary part of $\varepsilon_1$ (for a negative-permittivity material) to 0.23. Figure 2(c) shows the transmission curves corresponding to $\varepsilon_2$=3.5, 4.0, 4.3, 4.8 and 6.0, respectively, and Fig. 2(d) shows the corresponding distributions of magnetic field intensity at the image plane of the EHAL. For a guided mode with a large propagation constant $|k_y|$, the guided wave is highly localized around the waveguide. Consequently, the imaginary part of the propagation constant is very large when material loss exits. The embodiment of this point on the transmission curve of an EHAL is that the corresponding transmission peak disappears completely and the transmission $/t/$ of the evanescent waves around the peak is very small. On the contrast, the transmission peak corresponding to a guided mode with a small propagation constant $|k_y|$ is not completely damped by material loss and the transmission $/t/$ of the evanescent waves around the peak is still relatively large. As shown in Fig. 2(c), since the leftmost peaks for $\varepsilon_2$=4.0 and 4.3 are farther away from $k_0$ and the imaginary parts of their propagation constants are larger, the corresponding upheavals from the damped transmission peaks are lower and wider than that for $\varepsilon_2$=3.5. Even so, the upheavals on the transmission curves are still somewhat sharp, and the bottoms of the corresponding images are too extended as shown in Fig. 2(d). The image quality for the case of $\varepsilon_2$=4.8 is good (see curve 4 in Fig. 2(d)) since the evanescent waves in a large range are appropriately transmitted (see curve 4 in Fig. 2(c)).

In the above analysis, $d_1$=$d_2$=$a$/2 is fixed when a multilayered structure is homogenized. In fact, $d_1$ can be adjusted to other values, bigger or smaller than $a$/2. We have found, however, that if $d_1$ is not $a$/2, when the other parameters are fixed the value of $\varepsilon_2$ needs to be larger to get flat upheavals replacing sharp peaks on the transmission curve of the corresponding EHAL. Thus, we still fix $d_1$=$d_2$=$a$/2, which suffices to explain the key physics/thought of the suggested method. When $d_1$=$d_2$=$a$/2, $\varepsilon_2$ needs to be larger than $-\varepsilon_1$ ($\varepsilon_1$ is assumed to be real) in order to push the other transmission peaks toward the two peaks nearest $\pm k_0$ and obtain flat upheavals. Another reason for a larger $\varepsilon_2$ is as follows. If $\varepsilon_2$ is smaller than $-\varepsilon_1$ when $d_1$=$d_2$=$a$/2, one has $\varepsilon_x$>0 and $\varepsilon_y$<0 for the corresponding EHAL according to Eq. (1). The EFCs of the EHAL will then take the $y$ axis as the symmetric axis of parabolic curves instead of the $x$ axis for those in Fig. 2(b), and the incident propagation plane waves will be reflected strongly, which is not good for imaging. Because of the above reasons, the EHAL with $\varepsilon_x$=$\infty$ and $\varepsilon_y$=0 is investigated as a starting point to obtain high-subwavelength-resolution imaging. At last, another thing needs to be specified. If the thickness $d$ of an EHAL is very small compared with the wavelength $\lambda$, the value of $\varepsilon_2$ needs to be very large to obtain wide flat upheavals. At the infrared or visible frequencies, high-permittivity materials usually possess large material loss. And, since a lossy material absorbs the incident electromagnetic wave, the thickness $d$ can not be too large in order to obtain a bright image. Thus, moderate thickness $d$ ($d=9a=0.45 \lambda$) is used in the above discussion.

Appropriate interface termination is very important for subwavelength imaging. Appropriate interface termination has been used to generate surface waves to magnify incident evanescent waves and then increase the imaging solution of PCs [19]. Appropriate interface termination has also been used to enhance transmission [13,20,21]. In our previous work [22], we found that when the existence condition for the guided modes of a two-dimensional dielectric PC slab is just nearly fulfilled by adjusting the parameters of the interface layers, moderate amplification of evanescent waves in a wide range can be obtained. This thought is consistent with that in the previous part of this paper. Both rely on obtaining wide flat upheavals to transmit evanescent waves in a large range of $k_y$ appropriately. Below we use some coating layers as an improvement method to obtain high imaging resolution of multilayered structures. As an example, a coating layer with permittivity $\varepsilon_3 = 12.96$ and thickness $d_3=0.13a$ is put on one side of the EHAL used in Fig. 2(a). Other kinds of coating layers can also be used, which are not shown here. Figure 3(a) gives the corresponding transmission curves. Comparing Fig. 2(a) and Fig. 3(a), one can see that when the coating layer is added, smooth upheavals appear replacing the leftmost peaks on the transmission curves of the EHAL corresponding to $\varepsilon_2$=4.0 and 4.3, which means that the corresponding guided modes are not excited, but nearly excited. This procedure is not like the previously analyzed procedures where a smooth upheaval comes from merging two near peaks. The peaks with the smallest $|k_y|$ for $\varepsilon_2$=3.5 and 6.0 are so near $\pm k_0$ that they do not disappear, and therefore coating layers will not help for the cases of $\varepsilon_2$=3.5 and 6.0. Figure 3(b) shows the transmission curves when the imaginary part of $\varepsilon_1$ is set to 0.23, and Fig. 3(c) shows the corresponding distributions of magnetic field intensity at the image plane.

Although the image for $\varepsilon_2$=4.0 is improved, but the bottom is too extended, which indicates that the upheaval on curve 2 in Fig. 3(b) is still not enough flat. The image quality for $\varepsilon_2$=4.3 is improved greatly. Adding a coating layer is of less help for $\varepsilon_2$=4.8, for which the imaging quality has been good without a coating layer. In conclusion, additional coating layers can also improve the imaging quality.

## 3. NUMERICAL SIMULATION OF MULTILAYERED STRUTURES FOR HIGH-SUBWAVELENGTH-RESOLUTION IMAGING

In this section, subwavelength imaging of multilayered structures, which are not homogenized, is numerically analyzed with the transfer-matrix method. The EMT analysis in section 2 can give us some guidance, especially on the transmission for small $|k_y|$. The EMT comes from the second-order approximation analysis [23] and is therefore not accurate enough for large $|k_y|$. When loss exists, however, the transmission of large $|k_y|$ influences little the imaging of multilayered structures.

A lossless multilayered structure with $\varepsilon_1=-\varepsilon_2=-3.5$ and $d_1=d_2=a/2$ is first investigated. It satisfies $\varepsilon_x = \infty$ and $\varepsilon_y = 0$ when it is homogenized. Since the period $a$ of a multilayered structure can not be infinitely thin due to practical fabrication difficulty, $a=\lambda/20$ is used in calculation. The transmission curve of the multilayered structure is shown in Fig. 4(a) (curve 1). The EHAL with its parameters corresponding to curve 1 in Fig. 2(a) can be considered as a result from the homogenization of this multilayered structure. Comparing curve 1 in Fig. 2(a) and curve 1 in Fig. 4(a), one can see that for small $|k_y|$ the transmission $|t|$ of a multilayered structure can be predicted nearly well by the EHAL, including the positions of the transmission peaks with the smallest $|k_y|$. However, the transmission $|t|$ of the multilayered structure drops quickly for large $|k_y|$ and there appear many other peaks far away from $\pm k_0$, unlike that of the corresponding homogenized layer. These peaks representing guided modes come from the coupling of plasmon waves on the interfaces of the negative-permittivity layers, and more peaks appear when more periods are added [16]. When $\varepsilon_2$ increases from 3.5 to 4.0, 4.3, 4.8 and 6.0, the corresponding transmission curves are shown in Fig. 4(a), where we can see that the other peaks move toward the left and the leftmost peak moves toward the right. When $\varepsilon_2=4.8$, the leftmost two peaks merge together and a wide flat upheaval is generated. When $\varepsilon_2=6.0$, the smooth upheaval has disappeared and a new sharp peak appears very near $k_0$. The above phenomena occurring near $k_0$ can be predicted well by the analysis in section 2 since the EMT is valid for small $|k_y|$. The moving phenomenon of the other peaks (except the leftmost peak) toward the left is also predicted by the EMT, although their positions can not be predicted by the EMT when they are far away from $k_0$. Now we introduce some material loss and set the imaginary part of $\varepsilon_1$ to 0.23. The corresponding transmission curves are shown in Fig. 4(c) and the distributions of magnetic field intensity at the image plane are shown in Fig. 4(d) as $\varepsilon_2$ increases from 3.5 to 4.0, 4.3, 4.8 and 6.0. These transmission curves look quite similar to those shown in Fig. 2(c) for the corresponding EHALs. As shown in Fig. 4(c), the transmission $|t|$ for propagating plane waves is not low. Material loss enhances further the quick dropping of the transmission $|t|$ for large $|k_y|$ and these peaks with large $|k_y|$ disappear completely. As shown by transmission curve 4 for $\varepsilon_2=4.8$ in Fig. 4(c), the wide flat upheaval remains when loss exists. A high imaging resolution is obtained for this value of $\varepsilon_2=4.8$, and the full width at half maximum (FWHM) is about $0.114\lambda$ (see curve 4 in Fig. 4(d)).

Now we validate improving the subwavelength resolution of a multilayered structure by adding coating layers. As an example, a multilayered structure with $\varepsilon_1 = -3.5+0.23i$, $\varepsilon_2 = 4.0$, $d_1=d_2=a/2$ and $d=9a$ is investigated. The working wavelength is still $\lambda = 20a$. This multilayered structure is asymmetric. As illustrated in Fig. 1, the leftmost layer is a negative-permittivity layer and the rightmost layer is a dielectric layer. The imaging effects are completely different when a coating layer is added on the left side or the right side, unlike those for a symmetric EHAL. This reflects the limitation of the EMT from one aspect, but this does not impact the explanation of the mechanism of adding coating layers to improve subwavelength imaging with the EMT. When a coating layer with $\varepsilon_3 = 12.96$ and $d_3 = 0.13a$ is added on the right side of the multilayered structure, the transmission curve and imaging result are shown in Figs. 5(a) and 5(b) (curve 1), respectively. The imaging result is not improved at all. On the contrast, when the coating layer is added on the left side of the multilayered structure, a flat upheaval appears on curve 2 in Fig. 5(a). The imaging quality is improved greatly as shown by curve 2 in Fig. 5(b), and the FWHM is $0.096\lambda$. In addition, if the rightmost dielectric layer of the multilayered structure is also replaced with the same coating layer added on the left side, the imaging structure is now symmetric and the imaging result is shown in Fig. 5(b) (curve 3). Compared with the previous asymmetric case, the intensity and resolution (the FWHM is $0.095\lambda$) of the obtained image increase for the current case. In the above procedure, we coat layers of a third material, which may bring inconvenience for fabrication. Thus, sometimes it is preferable to coat layers of the same material as that of the dielectric layers or the negative-permittivity layers in a multilayered structure. Now a layer of the same material as the rightmost dielectric layer is added on the left side of the multilayered structure, and a symmetric structure is obtained. The corresponding transmission curve and imaging result are shown in Figs. 5(a) and 5(b) (curve 4), respectively.

The imaging quality also improves greatly. The image peak is very smooth and there are no side-lobes at all. The FWHM is about $0.103\lambda$, and the intensity of the image is low compared with that for adding coating layers of the third material since magnification of evanescent waves is further relaxed. On the contrast, if a layer of the same material as the leftmost negative-permittivity layer is added on the right side of the multilayered structure, over-magnification of evanescent waves occurs and a flat upheaval can not be obtained as shown by curve 5 in Fig. 5(a). The imaging quality becomes worse (curve 5 in Fig. 5(b)). In conclusion, when $\varepsilon_2$ is appropriately larger than $-\text{real}(\varepsilon_1)$, appropriate coating layers can usually improve subwavelength imaging.

All the multilayered structures we investigated before possess the same total thickness $d=9a$ when there are no coating layers. Here we note that when the other parameters are fixed, moderate change of $d$ does not change the imaging quality very significantly, unlike the symmetric multilayered structure investigated in [13] for which when a period is removed or added, the imaging quality becomes bad. This shows the advantage of our structure in fabrication tolerance. The multilayered structure (without any coating layer) corresponding to curve 4 of Fig. 4(a) is investigated as an example, which can realize high imaging resolution when material loss is considered. The parameters for the multilayered structure are $\varepsilon_1=-3.5$, $\varepsilon_2=4.8$, $d_1=d_2=a/2$, $a=\lambda/20$. Figure 6(a) shows the corresponding transmission curves when $d$ varies from $7a$ to $12a$. When $d=7a$, there is no flat upheaval on transmission curve 1 and two sharp peaks exist for evanescent waves with small $|k_y|$. When $d=8a$, the two sharp peaks merge and a smooth, but not enough flat, upheaval appears on transmission curve 2. When $d=9a$, a wide flat upheaval appears and this has been studied previously in this paper. As $d$ continues increasing, the upheaval begins to contract. At last when $d=12a$, the smooth upheaval disappears and a new peak is generated very near $k_0$. This phenomenon is similar to that when $\varepsilon_2$ increases with $d$ fixed. When the imaginary part of $\varepsilon_1$ is set to 0.23, Fig. 6(b) and 6(c) show the transmission and the imaging results corresponding to $d=7a$, …, and $12a$, respectively. For $d=7a$, there are two obvious high side-lobes and the imaging quality is bad. For $d=8a$, the imaging quality becomes better, but the two side-lobes are still obvious. The imaging quality for $d=9a$ is good. When $d$ increase further, the peak intensity becomes smaller and the resolution lower, but for $d=11a$ the FWHM is still as good as $0.129\,\lambda$. A similar result on fabrication tolerance can also be obtained for our multilayered structures with coating layers.

## 4. CONCLUSIONS

To obtain high-resolution imaging of a subwavelength object, it is important that an imaging structure should be able to transmit the evanescent waves from the object to the imaging plane with appropriate amplitudes in a large range. When a multilayered structure is used for subwavelength imaging, the guided modes with the smallest $|k_y|$ usually over-magnify the incident evanescent waves in some range and their amplitudes are not in appropriate proportion after transmitted to the image plane and consequently the imaging quality degrades. By choosing appropriate permittivity of the dielectric layers and adding a coating layer, we have solved such a problem and achieved high-subwavelength-resolution imaging. The transmission coefficients for propagating waves and evanescent waves are high and the obtained images are bright. In this paper, the negative permittivity is assumed to $\varepsilon_1=-3.5+0.23i$ as illustration. Since the suggested method is general, it can be applied for other negative permittivities and for various negative-permittivity materials, such as SiC, Ag, and Au, etc. For a noble metal, the imaginary part of the permittivity is larger compared with that of SiC, and typically the subwavelength imaging resolution is relatively lower. Only the case $d_1=d_2=a/2$ is investigated in this paper, but the suggested method can also be applied for other values of $d_1$ and $d_2$. The suggested method can also be applied to other kinds of subwavelength imaging systems, besides multilayered structures.

## ACKNOWLEDGEMENTS

This work is partly supported by the National Basic Research Program (No. 2004CB719801), the National Natural Science Foundations (NNSF) of China under Project No. 60688401 and Project No. 60677047, and the Swedish Research Council (VR) (No. 2006-4048).

## REFERENCES

1. J. B. Pendry, "Negative refraction makes a perfect lens," Phys. Rev. Lett. **85**, 3966-3969 (2000).

2. N. Fang, H. Lee, C. Sun, and X. Zhang, "Sub-Diffraction-Limited Optical Imaging with a Silver Superlens," Science **308**, 534–537 (2005).

3. D. Korobkin, Y. Urzhumov, and G. Shvets, "Enhanced near-field resolution in midinfrared using metamaterials," J. Opt. Soc. Am. B **23**, 468–478 (2005).

4. T. Taubner, D. Korobkin, Y. Urzhumov, G. Shvets, and R. Hillenbrand, "Near-Field Microscopy Through a SiC Superlens," Science **313**, 1595 (2006).

5. S. A. Ramakrishna, J. B. Pendry, M. C. K. Wiltshire, and W. J. Stewart, "Imaging the near field," J. Mod. Optics **50**, 1419–1430 (2003).

6. J. B. Pendry and S. A. Ramakrishna, "Refining the perfect lens," Phys. B **338**, 329–332 (2003).

7. P. A. Belov, Y. Hao, and S. Sudhakaran, "Subwavelength imaging at optical frequencies using a transmission device formed by a periodic layered metal-dielectric structure operating in the canalization regime," Phys. Rev. B **73**, 033108 (2006).

8. H. Shin and S. Fan, "All-angle negative refraction and evanescent wave amplification using one-dimensional metallodielectric photonic crystals," Appl. Phys. Lett. **89**, 151102 (2006).

9. B. Wood, J. B. Pendry, and D. P. Tsai, "Directed subwavelength imaging using a layered metal-dielectric system," Phys. Rev. B **74**, 115116 (2006).

10. K. J. Webb and M. Yang, "Subwavelength imaging with a multilayer silver film structure," Opt. Lett. **31**, 2130−2132 (2006).

11. S. Feng and J. M. Elson, "Diffraction-suppressed high-resolution imaging through metallodielectric nanofilms," Opt. Express **14**, 216−221 (2006).

12. M. Scalora, G. D'Aguanno, N. Mattiucci, M. J. Bloemer, D. de Ceglia, M. Centini, A. Mandatori, C. Sibilia, N. Akozbek, M. G. Cappeddu, M. Fowler, and J. W. Haus, "Negative refraction and sub-wavelength focusing in the visible range using transparent metallo-dielectric stacks," Opt. Express **15**, 508−523 (2007).

13. M. Bloemer, G. D'Aguanno, N. Mattiucci, M. Scalora, and N. Akozbek, "Broadband super-resolving lens with high transparency in the visible range," Appl. Phys. Lett. **90**, 174113 (2007).

14. C. Wang, Y. Zhao, D. Gan, C. Du, and X. Luo, "Subwavelength imaging with anisotropic structure comprising alternately layered metal and dielectric films," Opt. Express **16**, 4217−4227 (2008)

15. D. de Ceglia, M. A. Vincenti, M. G. Cappeddu, M. Centini, N. Akozbek, A. D'Orazio, J. W. Haus, M. J. Bloemer, and M. Scalora, "Tailoring metallodielectric structures for superresolution and superguiding applications in the visible and near-ir ranges," Phys. Rev. A **77**, 033848 (2008).

16. I. Avrutsky, I. Salakhutdinov, J. Elser, and V. Podolskiy, "Highly confined optical modes in nanoscale metal-dielectric multilayers," Phys. Rev. B **75**, 241402 (2007).

17. D. E. Aspnes, "Local-field effects and effective-medium theory: A microscopic perspective," Am. J. Phys. **50**, 704–709 (1982).

18. W. C. Chew, *Waves and Fields in Inhomogeneous Media*, 2nd ed. (IEEE Press, New York, 1995).

19. C. Y. Luo, S. G. Johnson, J. D. Joannopoulos, and J. B. Pendry, "Subwavelength imaging in photonic crystals," Phys. Rev. B **68**, 045115 (2003).

20. E. Moreno, F. J. García-Vidal, and L. Martín-Moreno, "Enhanced transmission and beaming of light via photonic crystal surface modes," Phys. Rev. B **69**, 121402 (2004).

21. Y. Jin and S. L. He, "Negative refraction of complex lattices of dielectric cylinders," Phys. Lett. A **360**, 461–466 (2007).

22. Y. Jin, X. Li, and S. L. He, "Canalization for subwavelength focusing by a slab of dielectric photonic crystal," Phys. Rev. B **75**, 195126 (2007).

23. J. Elser, V. A. Podolskiy, I. Salakhutdinov, and I. Avrutsky, "Nonlocal effects in effective-medium response of nanolayered metamaterials", Appl. Phys. Lett. **90**, 191109 (2007).

**Figure captions:**

Fig. 1. Multilayered structure composed of alternating negative-permittivity layers (dark) and positive-permittivity dielectric layers (light).

Fig. 2. (Color online) (a) Transmission curves of a lossless EHAL at $\lambda=20a$. The EHAL comes from homogenization of a multilayered structure with $\varepsilon_1=-3.5$, $d_1=d_2=a/2$, $d=9a$ and $\varepsilon_2$ increasing from 3.5 to 4.0, 4.3, 4.8 and 6.0. (b) EFCs at $\lambda=20a$ for the EHAL used in (a). Transmission curves (c) and distributions of magnetic field intensity at the image plane (d) of the EHAL used in (a) when the imaginary part of $\varepsilon_1$ is set to 0.23 with other parameters unchanged. In (a)–(d), curves 1–5 correspond to $\varepsilon_2=3.5$, 4.0, 4.3, 4.8 and 6.0, respectively.

Fig. 3. (Color online) (a) Transmission curves of a lossless EHAL with a coating layer added on one side at $\lambda=20a$. The EHAL comes from homogenization of a multilayered structure with $\varepsilon_1=-3.5$, $d_1=d_2=a/2$, $d=9a$ and $\varepsilon_2$ increasing from 3.5 to 4.0, 4.3, 4.8 and 6.0. The permittivity $\varepsilon_3$ and thickness $d_3$ of the coating layer are 12.96 and $0.13a$, respectively. Here we also show transmission curves (b) and distributions of magnetic field intensity at the image plane (c) of the EHAL with a coating layer used in (a) and the imaginary part of $\varepsilon_1$ set to 0.23. In (a)–(c), curves 1–5 correspond to $\varepsilon_2=3.5$, 4.0, 4.3, 4.8 and 6.0, respectively.

Fig. 4. (Color online) (a) Transmission curves of a multilayered structure with $d_1=d_2=a/2$, $d=9a$, $\varepsilon_1=-3.5$ and $\varepsilon_2$ increasing from 3.5 to 4.0, 4.3, 4.8 and 6.0 at $\lambda=20a$. (b) Local magnification of the part near $k_y=0$ in (a). Transmission curves (c) and distributions of magnetic field intensity at the image plane (d) when the imaginary part of $\varepsilon_1$ of the multilayered structure used in (a) is set to 0.23. In (a)–(d), curves 1–5 correspond to $\varepsilon_2$=3.5, 4.0, 4.3, 4.8 and 6.0, respectively. The results are accurately calculated with the transfer-matrix method.

Fig. 5. (Color online) (a) Transmission curves and distributions of magnetic field intensity at the image plane (b) when coating layers are added on one side or both sides of a multilayered structure with $\varepsilon_1$ =–3.5+0.23$i$, $\varepsilon_2$ =4.0, $d_1$=$d_2$=$a$/2 and $d$=9$a$ at $\lambda$ =20$a$. The multilayered structure having 9 periods is illustrated Fig. 1, with the leftmost layer a negative-permittivity layer and the rightmost layer a dielectric layer. Curve 1 (or 2) is for the case when a coating layer with $\varepsilon_3$ =12.96 and $d_3$=0.13$a$ is added on the right (or left) side of the multilayered structure, and curve 3 is for the case when a same coating layer is added on the left side and the other one replaces the rightmost dielectric layer simultaneously. Curve 4 (or 5) is for the case when a layer of the same material as the rightmost (or leftmost) dielectric (or negative-permittivity) layer is added on the left (or right) side of the multilayered structure. The results are accurately calculated with the transfer-matrix method.

Fig. 6. (Color online) (a) Transmission curves of a multilayered structure with $\varepsilon_1$=–3.5, $\varepsilon_2$=4.8, $d_1$=$d_2$=$a$/2 and $d$ varying from 7$a$ to 8$a$, …, and 12$a$ at $\lambda$=20$a$. Transmission curves (b) and distributions of magnetic field intensity at the image plane (c) of the multilayered structure used in (a) when the imaginary part of $\varepsilon_1$ is set to 0.23. In (a)–(c), curves 1–6 correspond to $d$=7$a$, 8$a$, …, and 12$a$, respectively. The results are accurately calculated with the transfer-matrix method.

**Figures:**

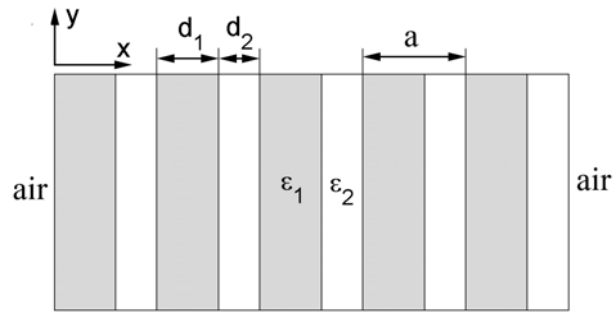

Fig. 1. Multilayered structure composed of alternating negative-permittivity layers (dark) and positive-permittivity dielectric layers (light).

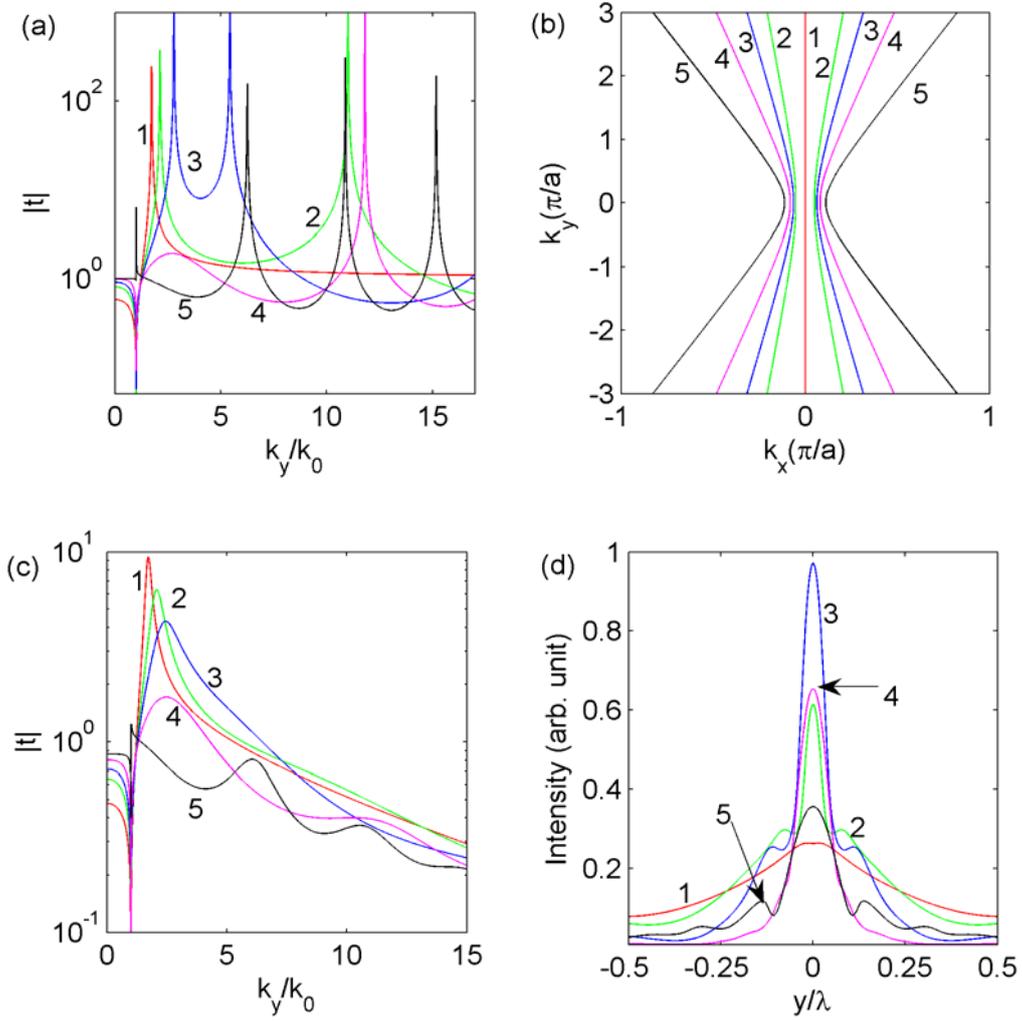

Fig. 2. (Color online) (a) Transmission curves of a lossless EHAL at $\lambda=20a$. The EHAL comes from homogenization of a multilayered structure with $\varepsilon_1=-3.5$, $d_1=d_2=a/2$, $d=9a$ and $\varepsilon_2$ increasing from 3.5 to 4.0, 4.3, 4.8 and 6.0. (b) EFCs at $\lambda=20a$ for the EHAL used in (a). Transmission curves (c) and distributions of magnetic field intensity at the image plane (d) of the EHAL used in (a) when the imaginary part of $\varepsilon_1$ is set to 0.23 with other parameters unchanged. In (a)–(d), curves 1–5 correspond to $\varepsilon_2=3.5$, 4.0, 4.3, 4.8 and 6.0, respectively.

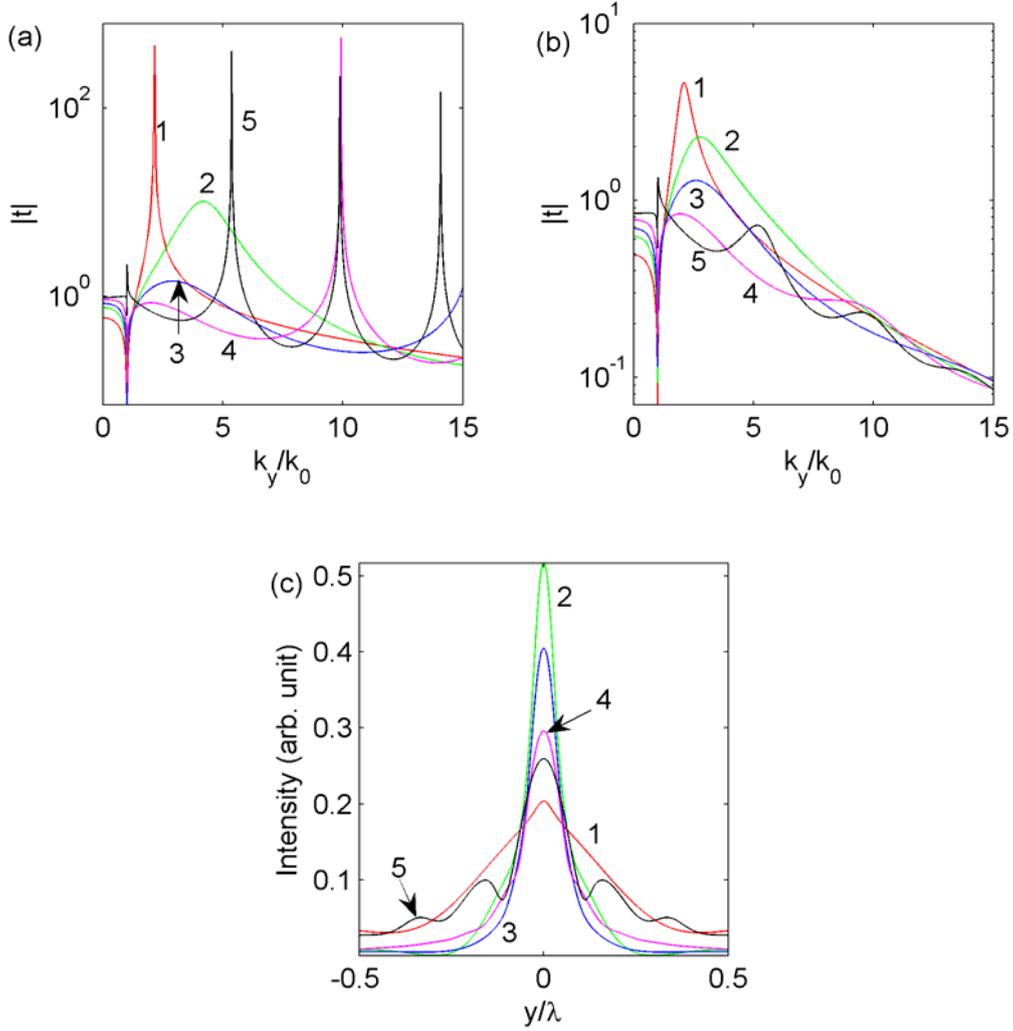

Fig. 3. (Color online) (a) Transmission curves of a lossless EHAL with a coating layer added on one side at $\lambda=20a$. The EHAL comes from homogenization of a multilayered structure with $\varepsilon_1=-3.5$, $d_1=d_2=a/2$, $d=9a$ and $\varepsilon_2$ increasing from 3.5 to 4.0, 4.3, 4.8 and 6.0. The permittivity $\varepsilon_3$ and thickness $d_3$ of the coating layer are 12.96 and $0.13a$, respectively. Here we also show transmission curves (b) and distributions of magnetic field intensity at the image plane (c) of the EHAL with a coating layer used in (a) and the imaginary part of $\varepsilon_1$ set to 0.23. In (a)–(c), curves 1–5 correspond to $\varepsilon_2=3.5$, 4.0, 4.3, 4.8 and 6.0, respectively.

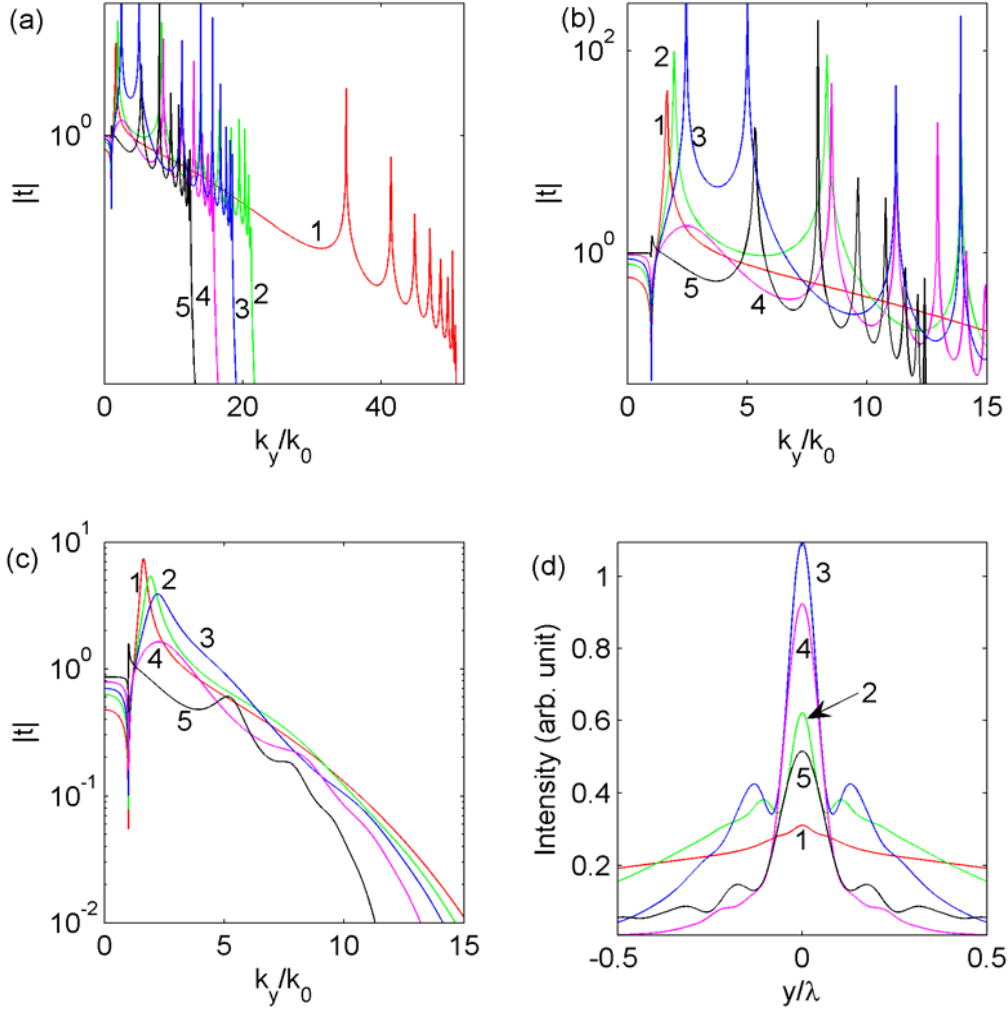

Fig. 4. (Color online) (a) Transmission curves of a multilayered structure with $d_1=d_2=a/2$, $d=9a$, $\varepsilon_1=-3.5$ and $\varepsilon_2$ increasing from 3.5 to 4.0, 4.3, 4.8 and 6.0 at $\lambda=20a$. (b) Local magnification of the part near $k_y=0$ in (a). Transmission curves (c) and distributions of magnetic field intensity at the image plane (d) when the imaginary part of $\varepsilon_1$ of the multilayered structure used in (a) is set to 0.23. In (a)–(d), curves 1–5 correspond to $\varepsilon_2=3.5$, 4.0, 4.3, 4.8 and 6.0, respectively. The results are accurately calculated with the transfer-matrix method.

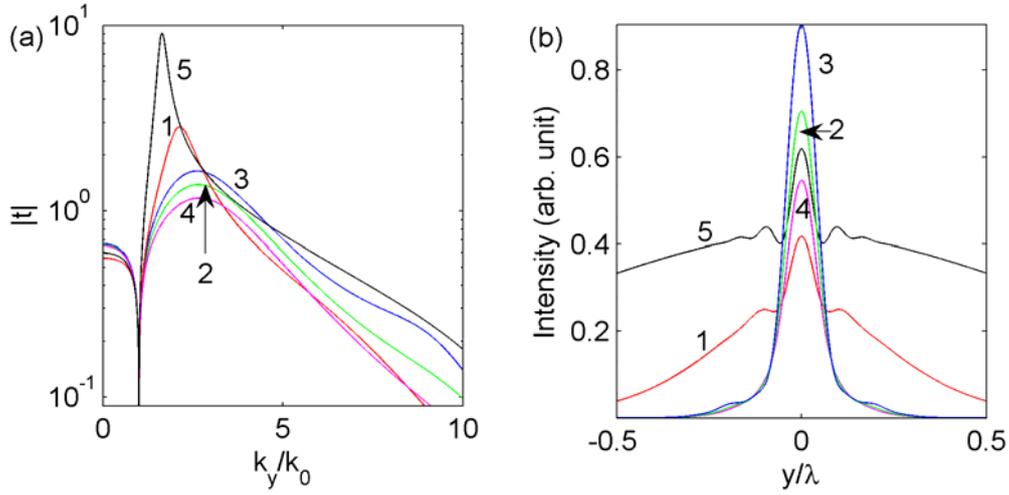

Fig. 5. (Color online) (a) Transmission curves and distributions of magnetic field intensity at the image plane (b) when coating layers are added on one side or both sides of a multilayered structure with $\varepsilon_1 = -3.5+0.23i$, $\varepsilon_2 = 4.0$, $d_1=d_2=a/2$ and $d=9a$ at $\lambda = 20a$. The multilayered structure having 9 periods is illustrated Fig. 1, with the leftmost layer a negative-permittivity layer and the rightmost layer a dielectric layer. Curve 1 (or 2) is for the case when a coating layer with $\varepsilon_3 = 12.96$ and $d_3 = 0.13a$ is added on the right (or left) side of the multilayered structure, and curve 3 is for the case when a same coating layer is added on the left side and the other one replaces the rightmost dielectric layer simultaneously. Curve 4 (or 5) is for the case when a layer of the same material as the rightmost (or leftmost) dielectric (or negative-permittivity) layer is added on the left (or right) side of the multilayered structure. The results are accurately calculated with the transfer-matrix method.

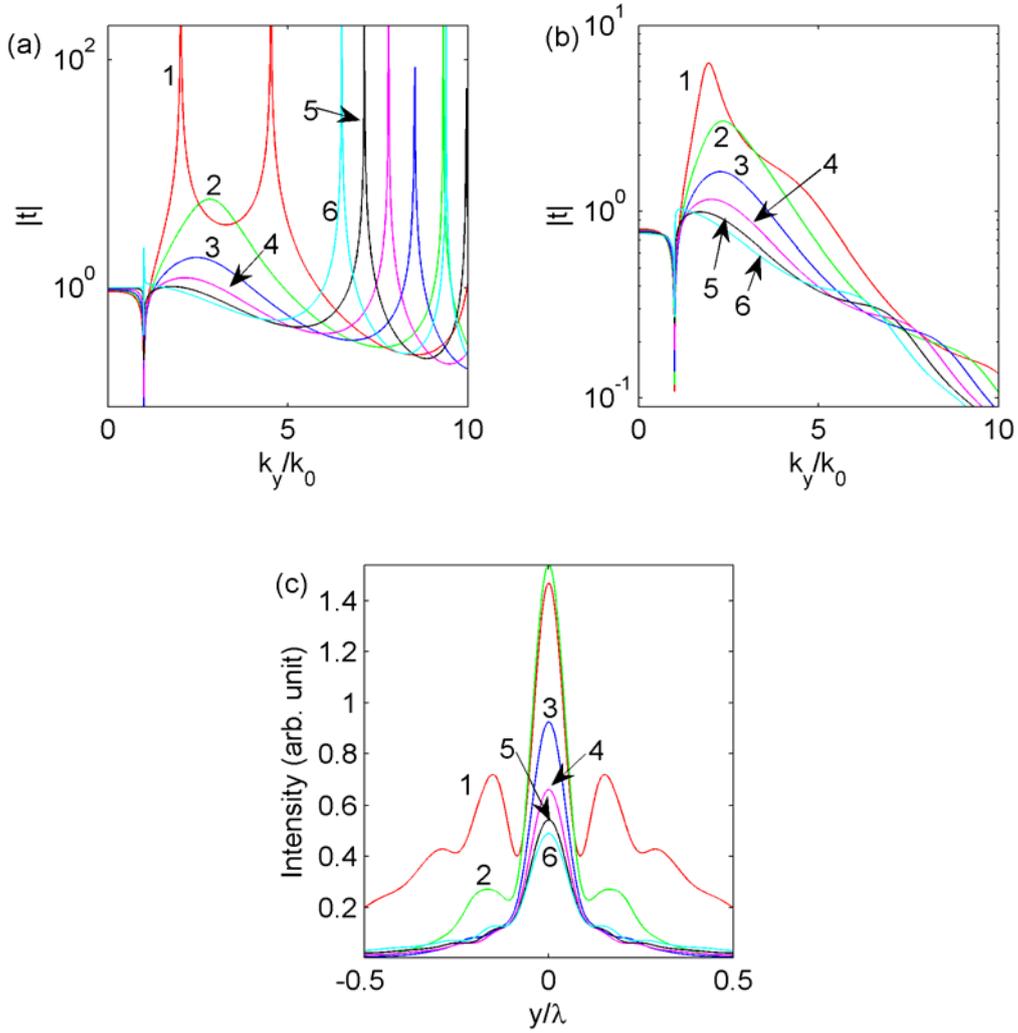

Fig. 6. (Color online) (a) Transmission curves of a multilayered structure with $\varepsilon_1=-3.5$, $\varepsilon_2=4.8$, $d_1=d_2=a/2$ and $d$ varying from $7a$ to $8a$, …, and $12a$ at $\lambda=20a$. Transmission curves (b) and distributions of magnetic field intensity at the image plane (c) of the multilayered structure used in (a) when the imaginary part of $\varepsilon_1$ is set to 0.23. In (a)–(c), curves 1–6 correspond to $d=7a$, $8a$, …, and $12a$, respectively. The results are accurately calculated with the transfer-matrix method.